\documentclass[prb,aps,showpacs,twocolumn,floatfix]{revtex4-1}
\usepackage{amsmath,amssymb,amsfonts,bm,graphicx}
\usepackage[english]{babel}
\newcommand{\bS}{{\bm S}}
\newcommand{\e}{{\rm e}}
\newcommand{\ii}{{\rm i}}

\newcommand{\vac}{\vert0\rangle}

\newcommand{\tr}{{\rm Tr}\,}
\newcommand{\str}{{\rm tr}}
\newcommand{\gd}{\gamma^\dag}

\newcommand{\Nbox}{{N_\Box}}
\begin{document}
\title{Frustrated magnets and quantum paramagnetic phases at finite
       temperature}
\author{L. Isaev$^1$}
\author{G. Ortiz$^2$}
\affiliation{$^1$Department of Physics and Astronomy,
                 Louisiana State University, Baton Rouge LA 80703 \\
	     $^2$Department of Physics and Center for Exploration of Energy and
	         Matter, Indiana University, Bloomington IN 47405}
\begin{abstract}
 We develop a general framework, which combines exact diagonalization in small
 clusters with a density matrix variational principle, to study frustrated
 magnets at finite temperature. This thermodynamic hierarchical mean-field
 technique is used to determine the phase diagram and magnetization process of
 the three-dimensional spin-$1/2$ $J_1$-$J_2$ antiferromagnet on a stacked
 square lattice. Its non-magnetic phase exhibits a thermal crossover from a
 quantum to a classical paramagnet at a temperature $T=T_0$ which can be
 extracted from thermodynamic measurements. At low temperature an applied
 magnetic field stabilizes, through order-by-disorder, a variety of phases with
 non-trivial spin textures and a magnetization plateau at half-saturation which
 continuously disappears at $T\sim T_0$. Our results are relevant for
 frustrated vanadium oxides.
\end{abstract}
\pacs{75.10.Jm, 75.10.Kt, 75.40.Cx, 75.60.Ej}
\maketitle

\paragraph*{Introduction.--}
Frustrated magnetic materials have been focus of active condensed matter
research in the past two decades \cite{Lacroix}. In these systems, competing
interactions and frustrated lattice topology often lead to fascinating effects,
e.g. magnetic monopoles \cite{ladak-2010,giblin-2011} in spin ice materials or
magnetization plateaux in an applied magnetic field \cite{Mila,isaev-2009a},
and stabilize exotic quantum states of matter, such as spin liquids
\cite{balents-2010} or valence bond solids \cite{matan-2010}. Understanding
these paramagnetic phases, which do not break any obvious global symmetry and
thus are not necessarily identified by an order parameter, is of fundamental
interest in material physics.

The theoretical description of frustration-driven phenomena is extremely
challenging, especially in spatial dimensions larger than one. For example,
frustrating interactions render large-scale quantum Monte-Carlo (QMC)
simulations impractical due to the ``sign problem'' \cite{Lauchli}. Various
approaches were developed to tackle frustrated magnets \cite{Richter}. One
class of methods proposes an expansion around a magnetically-ordered state
(spin-wave and series expansions), or in certain limiting cases, e.g. high
temperature. Another class focuses on ground state (GS) properties of the
system (coupled cluster and Lanczos methods). Meanwhile it is both
theoretically and experimentally \cite{balents-2010} relevant to inquire: What
are potential signatures of magnetic frustration in the thermodynamic
properties of a given material? Clearly, quantum effects due to a non-trivial
GS will become apparent at temperatures of the order of the characteristic
energy scales involved in the formation of that particular GS.

In the present Rapid Communication we address the above question. We develop an
unbiased and general framework aimed at studying the interplay between quantum
and thermal fluctuations in frustrated magnets. Our method couples the recently
developed hierarchical mean-field (HMF) theory \cite{isaev-2009} and the
well-known thermodynamic variational principle \cite{Ripka}. A key idea is the
realization that various competing {\it local} orders can only be captured
within an {\it exact diagonalization} scheme, while transitions between them
will be correctly described only in an {\it infinite} system. Hence, one starts
by partitioning a lattice into relatively small spin clusters (degrees of
freedom) in accordance with point-group symmetries. These new degrees of
freedom provide a language in which the original model Hamiltonian is
represented. Approximations are introduced in the form of a variational
principle applied to the free energy in the new representation. Our approach
relies on numerical as well as analytical efforts: provided the cluster is
chosen properly, even a simple variational ansatz for the density matrix (DM)
yields a {\it complete} phase diagram of the system. Results can be
systematically improved by considering larger clusters or more complicated
trial DMs. Once the system DM is known, any observable can be computed even
{\it inside} non-magnetic (paramagnetic) phases, in contrast to the usual
mean-field (MF) techniques which only yield instabilities of the
magnetically-ordered states.

We illustrate the thermodynamic HMF (THMF) formalism by studying the phase
diagram at finite temperature $T$ and properties in an applied magnetic field
of the ``stacked'' $J_1$-$J_2$ model \cite{richter-2006,nunes-2011,rojas-2011}
which describes a spin-$1/2$ Heisenberg antiferromagnet on an orthorhombic
lattice with first ($J_1$) and second ($J_2$) neighbor interactions in the
$ab$-plane, and a nearest-neighbor (NN) exchange ($J_C$) along the $c$-axis.
This model is a good approximation for layered vanadium oxide materials
\cite{thalmeier-2009,thalmeier-2008,schmidt-2009}, such as
${\rm Li_2VO(Si,Ge)O_4}$ \cite{bombardi-2004} and ${\rm PbVO_3}$
\cite{tsirlin-2008}. Previous works indicate that at $T=0$ the
$J_1$-$J_2$-$J_C$ model exhibits a quantum paramagnetic phase whose stability
can be controlled by changing $J_C$. We show that this phase persists even at
finite $T$ and introduce an important temperature scale, $T_0$, at which a
crossover from quantum to classical paramagnetic behavior takes place.
Numerical value of $T_0$ can be extracted from thermodynamic or magnetization
measurements. Below $T_0$, an interplay between quantum fluctuations inside the
paramagnetic region and external field results in a variety of phases
characterized by non-trivial magnetic orders, and leads to a magnetization
plateau around half of the saturation field. Our findings are directly relevant
for the frustrated perovskite ${\rm PbVO_3}$ which shows no magnetic order and
is believed to realize a $J_1$-$J_2$ quantum paramagnet \cite{tsirlin-2008}.

\paragraph*{The {\rm THMF} method.--}
Let us consider a quantum spin model defined on a lattice with $N$ sites and
periodic boundary conditions. Following the HMF prescription \cite{isaev-2009},
we partition the lattice into $\Nbox$ clusters of $N_q$ sites each, so that
$N=\Nbox N_q$. It is assumed that eigenstates of an isolated cluster are known.
Each cluster state $\vert a\rangle$ can be associated with a Schwinger boson
$\gd_a$, subject to a constraint $\sum_a\gd_a\gamma_a=1$. It is important to
note that the version of THMF method developed below treats this constraint
{\it exactly}, a condition that is crucial for the method to be variational.

In terms of ``cluster'' degrees of freedom the original Hamiltonian of the
system can be written as:
\begin{equation}
 H=\sum_i(H_0)_{ab}\gd_{ia}\gamma_{ib}+
 \sum_{ij}(V_{ij})^{a^\prime b^\prime}_{ab}\gd_{ia^\prime}
 \gd_{jb^\prime}\gamma_{jb}\gamma_{ia},
 \label{cluster_hamiltonian}
\end{equation}
with $H_0$ and $V$ being the cluster self-energy and inter-cluster
interaction, respectively. The subscript $i$ labels different clusters in the
coarse-grained lattice and summation over doubly repeated indices is assumed.
The range and type of couplings in the second term are determined by the
original model. Each spin (and therefore each cluster) has $\nu_\lambda$ links
of type-$\lambda$ ($\lambda=1$, $2$, $\ldots$) with corresponding interactions
$V_\lambda$. For instance, on a square $J_1$-$J_2$ lattice there are $\nu_1=2$
NN and $\nu_2=2$ next-NN (NNN) links per site. The Hamiltonian
\eqref{cluster_hamiltonian} operates on a Hilbert space spanned by the products
$\vert\lbrace a\rbrace\rangle=\prod_i\gd_{ia_i}\vac$, where $\vac$ is the
unphysical Schwinger boson vacuum corresponding to all ``empty'' clusters.

The THMF theory is a variational approach with respect to the free energy. Here
we consider a simple trial DM $\rho_0\bigl[H_{\rm MF}\bigr]=\e^{-K/T}/Z_0$ with
$K=\sum_i\bigl(H_{\rm MF}\bigr)_{ab}\gd_{ia}\gamma_{ib}$, where the Boltzmann
constant $k_B\equiv1$, and the MF Hamiltonian $H_{\rm MF}$, whose matrix
elements play the role of variational parameters \cite{Ripka}, is
self-consistently determined by minimizing the free energy
$F\!=\!E-TS=\tr\rho_0(H+T\ln\rho_0)=\tr\rho_0(H-K)-T\ln Z_0$. The partition
function can be expressed in terms of the eigenvalues $E_n$ of $H_{\rm MF}$ as
$Z_0=Z_1^\Nbox$ with $Z_1=\sum_n\e^{-E_n/T}$. Then the free energy becomes:
\begin{displaymath}
 \frac{F}{\Nbox}=-T\log Z_1+\str_1\omega_1\biggl[H_0+\sum_\lambda\nu_\lambda
 \str_2(V_\lambda)_{12}\omega_2-H_{\rm MF}\biggr].
\end{displaymath}
Here $\omega=\e^{-H_{\rm MF}/T}/Z_1$ is the single-cluster DM, ``$\str$''
denotes a trace over single-cluster MF states, and
$\str_1\str_2(V_\lambda)_{12}\omega_1\omega_2=
(V_\lambda)^{a_1^\prime a_2^\prime}_{a_1a_2}\omega_{a_1a_1^\prime}
\omega_{a_2a_2^\prime}$.

A simple calculation yields the MF Hamiltonian \cite{Ripka}:
\begin{equation}
 (H_{\rm MF})_{a^\prime a}=(H_0)_{a^\prime a}+2\sum_\lambda\nu_\lambda
 (V_\lambda)^{a^\prime b_1}_{ab_2}\omega_{b_2b_1}.
 \label{hmf_hamiltonian}
\end{equation}
Using its eigenstates $R_a^n$ one can compute the DM $\omega$:
\begin{equation}
 \omega_{ab}=\sum_n\e^{-E_n/T}R_a^n(R_b^n)^*\biggl /\sum_n\e^{-E_n/T}
 \label{single_cluster_rho}
\end{equation}
and the free energy $F/N=(1/2N_q)\tr\omega(H_0-H_{\rm MF})-(T/N_q)\ln Z_1$.
In the limit $T\to0$ only the GS eigenpair $(E_0,R^0)$ contributes to the DM
\eqref{single_cluster_rho}, i.e. $\omega_{ab}\to R^0_a(R^0_b)^*$, and we
recover the $T=0$ HMF results \cite{isaev-2009}.

\begin{figure}[t]
 \begin{center}
  \includegraphics[width=0.7\columnwidth]{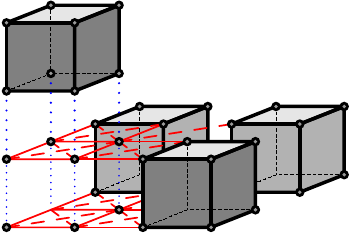}
 \end{center}
 \caption{The $J_1$-$J_2$-$J_C$ lattice. Solid, dashed and dotted lines
          denote $J_1$, $J_2$ and $J_C$ exchange interactions respectively.
          Shaded cubes are clusters used in the THMF calculation.}
 \label{fig_lattice}
 \vspace{-0.5cm}
\end{figure}

Expressions \eqref{hmf_hamiltonian} and \eqref{single_cluster_rho} define the
THMF self-consistent (in terms of $\omega_{ab}$) scheme. Any local observable
$\langle{\cal O}\rangle$ can be computed as
$\langle{\cal O}\rangle=\str\,\omega{\cal O}$.

\paragraph*{$J_1$-$J_2$-$J_C$ model.--}
We now use the THMF method to study the $J_1$-$J_2$-$J_C$ model
\cite{richter-2006} (Fig. \ref{fig_lattice}):
\begin{equation}
 H=\Biggl(J_1\sum_{\langle ij\rangle}+
 J_2\sum_{\langle\langle ij\rangle\rangle}+J_C\sum_{\langle ij\rangle_z}\Biggr)
 \bS_i\bS_j-h\sum_iS^z_i,
 \label{j1_j2_jc}
\end{equation}
where all couplings $J_{1,2,C}$ are positive, $\bS_i$ is a spin-$1/2$
operator at site $i$ and $h$ is an external magnetic field. This model
describes a three-dimensional (3D) analog of the $J_1$-$J_2$ antiferromagnet
\cite{Misguich} defined on a cubic lattice with intralayer NN
($\langle ij\rangle$) and NNN ($\langle\langle ij\rangle\rangle$) interactions
$J_1$ and $J_2$ respectively, and an interlayer NN ($\langle ij\rangle_z$)
exchange $J_C$. In the following we adopt the units $J_1\equiv1$.

At $h=T=0$, the coupling $J_C$ can be viewed as a tuning parameter controlling
quantum effects associated with frustration. In particular, for $J_C>J_C^0$ the
quantum paramagnetic phase of the $J_1$-$J_2$ model disappears and the system
exhibits a direct 1st order transition from N\'eel to a columnar
antiferromagnetic (AF) state. For $J_C<J_C^0$ there is an intermediate
non-magnetic region which appears for a finite range of $J_2$ and vanishes at
$J_C^0$. Thus the point $[J_2(J_C^0),J_C^0]$ is a multicritical point where
three phase boundaries converge. There seems to be a controversy regarding the
order of phase transitions occurring at these boundaries: spin-wave studies
\cite{nunes-2011} predict one 2nd and two 1st order lines, while series
expansions \cite{rojas-2011} indicate that all phase boundaries are 1st order.
Our analysis supports the spin-wave scenario \cite{dqc}.

We apply the THMF theory to compute the finite-temperature phase diagram of
Hamiltonian \eqref{j1_j2_jc}. While the effect of thermal fluctuations on
ordered phases is well known, their role inside a {\it quantum} non-magnetic
state is unclear. Since this state does not break any continuous symmetry, the
system can only undergo a crossover from a quantum to a standard classical
paramagnet. We will show how the corresponding temperature scale can be deduced
from experimentally accessible thermodynamic quantities, such as specific heat
or uniform susceptibility. We also study properties of the $J_1$-$J_2$-$J_C$
model in an applied magnetic field. In the paramagnetic phase the magnetization
curve has a plateau, similar to that of Ref. \onlinecite{zhitomirsky-2000}
obtained for the two-dimensional (2D) $J_1$-$J_2$ model at $T=0$. We
demonstrate that with increasing field, a system exhibits a set of metamagnetic
transitions characterized by non-coplanar spin textures.

\begin{figure}[t]
 \begin{center}
  \includegraphics[width=0.9\columnwidth]{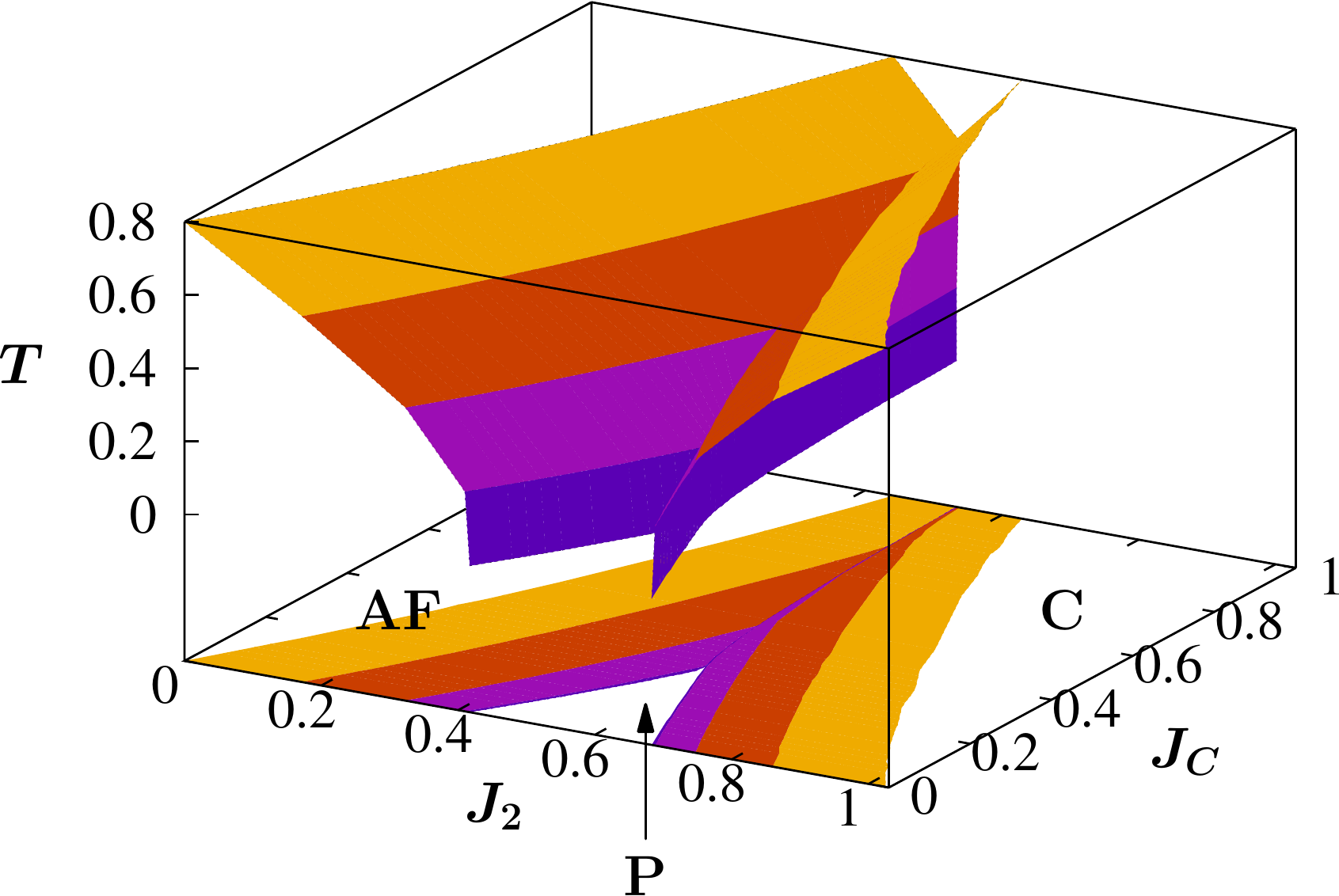}
 \end{center}
 \caption{Phase diagram of the $J_1$-$J_2$-$J_C$ model at $h=0$ ($J_1=1$).
          There are three phases: $(\pi,\pi,\pi)$-N\'eel antiferromagnet (AF),
          $(0,\pi,\pi)$-columnar AF (C), and paramagnet (P).}
 \label{fig_phase_diagram}
 \vspace{-0.5cm}
\end{figure}

Application of the THMF technique starts by selecting a proper degree of
freedom. Since the $J_1$-$J_2$-$J_C$ lattice has an orthorhombic unit cell with
symmetry $D_{2h}$ \cite{Bir}, the smallest cluster compatible with this
symmetry is a cube, shown in Fig. \ref{fig_lattice}. Each cluster has $6$ NN
($4$ in the $ab$-plane, $2$ along the $c$-axis) and $4$ NNN clusters in the
$ab$-plane; i.e. $\nu_1=\nu_2=2$ and $\nu_C=1$ in \eqref{hmf_hamiltonian}.
Next, we compute the matrices $H_0$ and $V$ in Eqs. \eqref{cluster_hamiltonian}
and \eqref{hmf_hamiltonian} using the method of Ref. \onlinecite{isaev-2009}.
Finally, $\omega_{ab}$ is determined self-consistently using Eqs.
\eqref{hmf_hamiltonian} and \eqref{single_cluster_rho}.

\paragraph*{Finite-T phase diagram.--}
The zero-field phase diagram of the $J_1$-$J_2$-$J_C$ model \eqref{j1_j2_jc} is
presented in Fig. \ref{fig_phase_diagram}. At $T=0$ and $J_C<J_C^0$ there exist
three phases: AF with the wavevector ${\bm Q}=(\pi,\pi,\pi)$, plaquette quantum
paramagnet (P), and columnar AF (C) with ${\bm Q}=(0,\pi,\pi)$ or
$(\pi,0,\pi)$. The magnetic phases are characterized by an order parameter of
the form $M_i=\langle\bS_i\rangle=M_{\bm Q}\e^{\ii{\bm Q}{\bm x}_i}$. The
transition AF-P (P-C) is 2nd (1st) order for all values of $J_C$. This
conclusion is consistent with the phase diagram of the $J_1$-$J_2$ ($J_C=0$)
model \cite{isaev-2009} and agrees with the spin-wave \cite{nunes-2011} and
coupled cluster \cite{richter-2006} analysis. For $J_C>J_C^0$ there is only a
1st order AF-C transition. The THMF calculation yields
$J^0_C(T=0)\sim0.28-0.30$, in agreement with previous works
\cite{richter-2006,nunes-2011,rojas-2011}. The real-space structure inside the
P-region is {\it plaquette crystal-like} with layers covered with $2\times2$
plaquettes, each being in its singlet GS. This state is similar to the
paramagnetic GS of the 2D $J_1$-$J_2$ model \cite{isaev-2009} and is stabilized
because for the P-phase $J_C$ is quite small, so the system exhibits quasi-2D
behavior.

\begin{figure}[t]
 \includegraphics[width=\columnwidth]{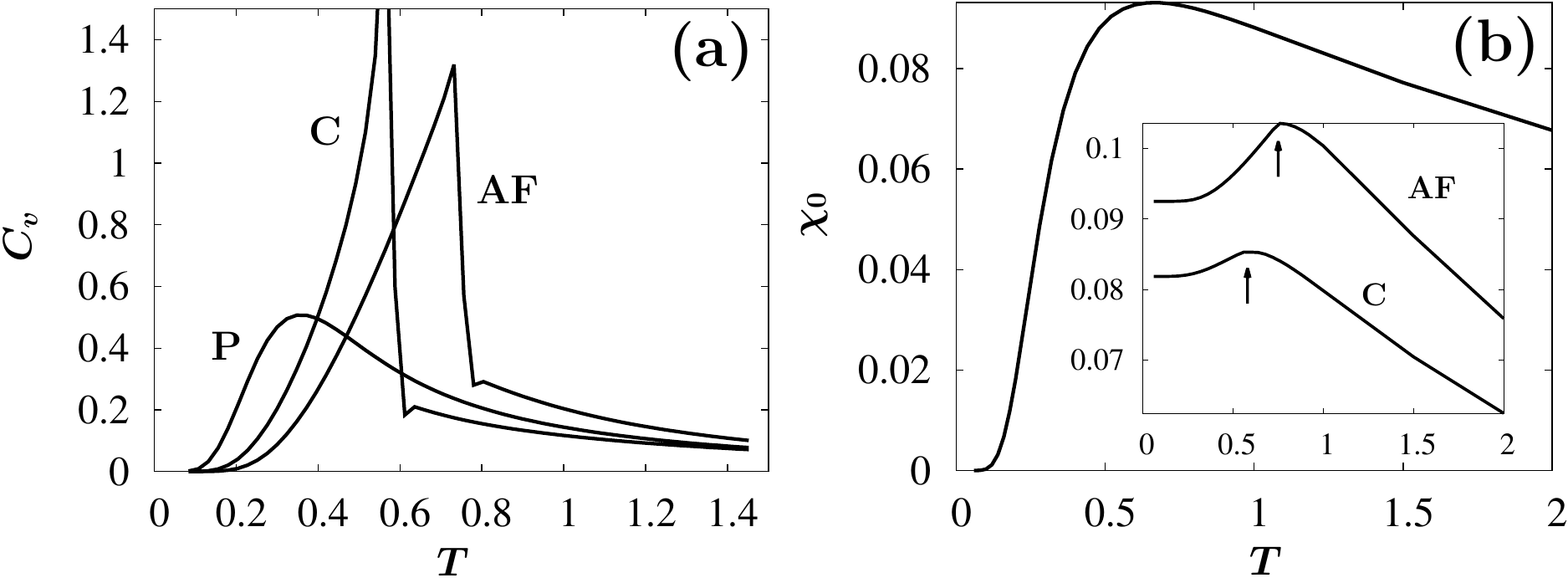}
 \caption{(a) Specific heat $C_v$ for the phases in Fig.
          \ref{fig_phase_diagram} at $J_C=0.1$: AF ($J_2=0.1$), P ($J_2=0.5$)
          and C ($J_2=0.8$). (b) Uniform susceptibility $\chi_0$ for the
          paramagnetic phase (main plot), and magnetic states (inset) with
          arrows indicating the N\'eel temperature. Parameters are the same as
          in panel (a).}
 \label{fig_Cv_chi0}
 \vspace{-0.5cm}
\end{figure}

With increasing $T$ the ordered AF- and C-states are suppressed via a 2nd order
(classical) phase transition at a N\'eel temperature $T_N(J_2,J_C)$, which is
accompanied by a jump in the magnetic specific heat $C_v$, see Fig.
\ref{fig_Cv_chi0}(a). Since the THMF method ignores long-range fluctuations,
$T_N$ remains finite even in the 2D limit $J_C=0$, which constitutes a
violation of the Mermin-Wagner theorem \cite{mermin-1966} common to MF
theories. We note that in the C-state along with the columnar magnetization
$M_c$ one can introduce an Ising order parameter
\cite{chandra-1990,richter-2008}
$\sigma=(1/N)\sum_{x,y}{\bm S}_{x,y}\bigl({\bm S}_{x+1,y}-{\bm S}_{x,y+1}
\bigr)$, with the summation extending over $N$ sites of the original lattice
and $(x,y)\equiv i$. This $\mathbb{Z}_2$ order parameter is not subject to the
conditions of the Mermin-Wagner theorem and vanishes at a non-zero temperature
\cite{chandra-1990} via a 2nd order phase transition. However, at the HMF level
$\sigma$ and $M_c$ vanish at the same N\'eel temperature.

Inside the P-phase, a crossover takes place from quantum (at low $T$) to
classical (at high $T$) behavior. Because the paramagnetic GS is gapped, $C_v$
exhibits a peak [Fig. \ref{fig_Cv_chi0}(a)]. We argue that the position of this
peak serves as an estimate of the crossover temperature $T_0$ at which thermal
fluctuations become comparable to the gap in the GS. For instance, at $J_2=0.5$
and $J_C=0.1$, $T_0\sim0.3$. The gap in the P-state manifests itself in an
activated behavior of the uniform linear magnetic susceptibility
$\chi_0(T)=\partial M_z/\partial h\vert_{h=0}$, presented in Fig.
\ref{fig_Cv_chi0}(b). Contrary to the ordered AF- and C-states where
$\chi_0(T=0)$ is finite and has a peak approximately at the corresponding
$T_N$, for the P-phase $\chi_0(T=0)=0$ and shows a broad maximum around $T_0$.
Since the P-phase has a plaquette structure, it is natural to examine the
behavior of various plaquette ``order parameters''. We considered the functions
$F_4=(1/N)\sum_{x,y}{\bm S}_{x,y}\bigl[(-1)^x{\bm S}_{x+1,y}+
(-1)^y{\bm S}_{x,y+1}\bigr]$ and $Q=(1/2N)\sum(P_{1234}+P^{-1}_{1234})$
introduced in Refs. \onlinecite{richter-2008} and \onlinecite{fouet-2003},
respectively. In the expression for $Q$ the summation takes place over
plaquettes in $ab$-planes and $P_{1234}$ is an operator of cyclic permutation
of plaquette vertices. Both functions remain finite and decay as $\sim1/T$ at
large temperature. The crossover manifests itself through a peak in $dF_4/dT$
and $dQ/dT$ around $T_0$. As any real-space method, the THMF theory involves
explicit translational symmetry breaking (cf. Ref. \onlinecite{dqc}),
predicting a crossover inside the paramagnetic phase. In an exact thermodynamic
limit solution this crossover may become a phase transition because of melting
of the plaquette crystal.

Now let us consider transitions between different phases, which are triggered
by tuning $J_2$ while keeping $T>0$ and $J_C$ fixed. The transition AF-P
remains 2nd order. On the other hand, the transition P-C is clearly
discontinuous at low $T$ because symmetries of the two phases are not related
by the group-subgroup relation \cite{isaev-2009}. At higher $T$ (still not
destroying the magnetic order), the jump in the columnar magnetization vanishes
and the transition becomes continuous.

Finally we compare the critical temperature and exponents obtained from the
8-spin cluster THMF with their known values. For the unfrustrated 3D Heisenberg
model ($J_2=0$ and $J_C=1$) the specific heat and order parameter exponents are
$\alpha_{THMF}=0$ and $\beta_{THMF}=0.470(1)$, and the N\'eel temperature is
$T_N^{THMF}=1.308$. These values should be compared with the results of the
Weiss molecular field $\alpha_W=0$, $\beta_W=0.5$, $T_N^W=1.5$, and QMC
\cite{holm-1993,sengupta-2003} $\beta_{QMC}=0.36$, $T_N^{QMC}=0.946$. Near the
paramagnetic phase, e.g. for $J_2=0.3$ and $J_C=0.1$, we have
$\beta_{THMF}=0.474(2)$ and $T_N^{THMF}=0.512$. The THMF exponents and $T_N$
can be improved by increasing the cluster size and implementing a finite-size
extrapolation.

\begin{figure}[t]
 \begin{center}
  \includegraphics[width=\columnwidth]{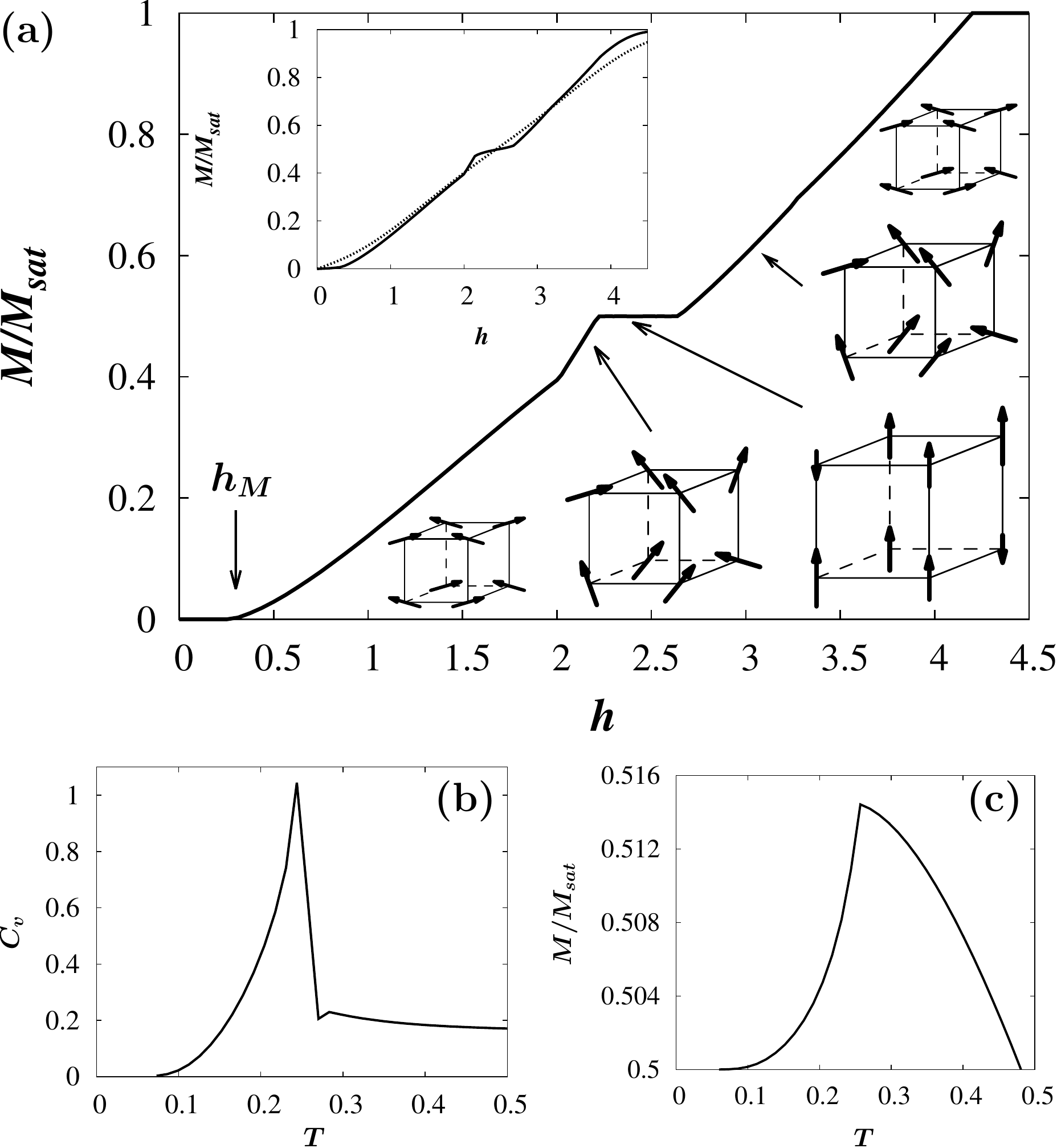}
 \end{center}
 \caption{(a) Magnetization process of the $J_1$-$J_2$-$J_C$ model with
	  $J_2=0.5$, $J_C=0.1$, and $M_{sat}=1/2$. Main panel: Total
	  magnetization $M$ versus applied field at $T=0$ and schematic spin
	  profiles in the corresponding field ranges (the length of the arrows
	  is proportional to the magnitude of the spin expectation values).
	  Inset: $M$ at finite temperature $T=0.16$ (solid line) and $T=0.3$
	  (dotted line). (b) and (c) Temperature dependence of the specific
	  heat and $M$ at $h=2.5$. The 2nd order phase transition happens at
	  $T\sim T_0$.}
 \label{fig_mag}
 \vspace{-0.5cm}
\end{figure}

\paragraph*{Magnetic states in an applied field.--}
An exhaustive spin-wave study of the magnetization process $M(h)$ inside
ordered phases of the $J_1$-$J_2$-$J_C$ model was performed in Ref.
\onlinecite{thalmeier-2008}. On the contrary, high-field properties of the
quantum paramagnetic state received much less attention, with efforts
exclusively focused on the case $J_C=0$ (2D $J_1$-$J_2$ model). Particularly,
in Ref. \onlinecite{zhitomirsky-2000} a half-saturation ($M=1/4$) magnetization
plateau characterized by a collinear spin ordering was proposed around the
maximally frustrated point $J_2=0.5$.

Here we apply the THMF theory to study field-induced metamagnetic transitions
inside the non-magnetic region (Fig. \ref{fig_phase_diagram}) of the
$J_1$-$J_2$-$J_C$ model with $J_2=0.5$ and $J_C=0.1$. The main panel of Fig.
\ref{fig_mag}(a) displays the $T=0$ magnetization curve. For $h$ below certain
threshold value $h_M$, the magnetization vanishes due to the spectral gap. For
$h>h_M$, the system exhibits a set of phases with non-coplanar magnetic
textures, shown in the figure. While the spin orderings at small and large
fields simply reflect canted AF sublattices, the structures immediately before
and after the $1/2$-plateau are non-trivial because the classical canting angle
varies for different spins. The plateau state has long-range order
characterized by a collinear magnetic structure with two spins per cluster
antiparallel to the field (cf. Ref. \onlinecite{zhitomirsky-2000}) and an Ising
order parameter.

The plateau width [Fig. \ref{fig_mag}(a), inset] and threshold field $h_M$
vanish at $T\sim T_0$ in agreement with the behavior of $C_v$ [Fig.
\ref{fig_Cv_chi0}(a)]. For a fixed field inside the plateau, the magnetic order
collapses via a 2nd order phase transition at $T\sim T_0$. In Fig.
\ref{fig_mag}(b) and (c) we show the temperature dependence of $C_v$ and $M$ at
$h=2.5$. The non-monotonic behavior of $M(T)$ is easy to understand by
observing that at $T=0$ spins are ``locked'' in a specific pattern by the
interactions. Thermal fluctuations unlock the spins allowing them to orient
along the field, thus increasing $M(T)$ before the transition. This highly
non-trivial magnetization process is specific to the P-state in Fig.
\ref{fig_phase_diagram}, and we propose to use it in conjunction with the peak
in $C_v$ as characteristic signatures of a quantum paramagnet.

\paragraph*{Conclusion.--}
We developed and applied the THMF method to address the interplay between
quantum and thermal fluctuations in the spin-$1/2$ $J_1$-$J_2$-$J_C$
antiferromagnet. Focusing on the non-magnetic region of the model, which is
inaccessible for other theoretical techniques, we studied the crossover between
a classical (due to thermal effects) and a quantum paramagnet, and demonstrated
how the crossover temperature scale $T_0$ can be extracted from thermodynamic
and high-field measurements. At low temperature $T<T_0$ quantum fluctuations
inside the paramagnetic state are manifested in a variety of field-induced spin
structures and a magnetization plateau at half-saturation. Our results can be
verified in experiments with vanadium oxides of the type ${\rm PbVO_3}$.
Assuming \cite{tsirlin-2008} that $J_1\sim70-100$~K (and $J_2$, $J_C$ inside
the paramagnetic phase), one gets $T_0\sim20-30$~K and $h_M\sim25-35$~T. The
magnetization plateau should become apparent for $h\sim100-200$~T.

\paragraph*{Acknowledgments.--}
We thank J. Dukelsky for illuminating discussions. L.I. was supported by DOE
via Grant DE-FG02-08ER46492.

\end{document}